# Research on a new active power filter topology based on chopper circuit

Guo Xiaoling(郭晓玲)[1,2], Cheng Jian(程健)[3], Zhang Bo(张波)[4], Zhang Jing(张旌)[1,2,3]

[1] Dongguan Campus, Institute of High Energy Physics, Chinese Academy of Sciences (CAS), Dongguan 523803, China

[2] Dongguan Institute of Neutron Science, Dongguan, 523808, China

[3] Institute of High Energy Physics, Chinese Academy of Sciences, Beijing 100049, China

[4] BoXingKeYuan Electronic Technology Co., Ltd., Beijing, China, 1013000

**Abstract:** The active power filter (APF) is attracting more and more attention for its outstanding performance in current and voltage ripple compensation. As modern high-energy accelerators are demanding much more stringent current ripple guideline, the APF is introduced to the magnet power supply (MPS) in accelerator system. However, the conventional APF has a lot of shortages and drawbacks due to its traditional topology, such as complex structure, nonadjustable working voltage, requirement of power supply, and so on. This paper proposes a new topology of APF, which is working as two types of chopper circuits. This APF doesn't need extra electricity, but to use the power of the MPS current ripple to realize ripple depressing. At the end of this paper, the experiment result proves its feasibility and effect.

## 1. Introduction

In MPS system, the APF can effectively depress the ripples or harmonics in current and voltage, thus improving the current's quality and stability in accelerator. To compensate the ripple component in MPS, the APF outputs a current to the magnet, which has the same amplitude but opposite phase with the MPS current ripple, thereby counteracts the current ripple through the magnet [1].

There are two types of APFs, which are parallel-APF and series-APF. Between them, the parallel APF is used to compensate the current ripple, while the series one is used to decrease the voltage ripple. As most magnet power supplies are high-precision current source, the parallel-APF is more suitable and appropriate in accelerator MPS. Therefore this paper only focuses on the parallel one[2].

Obviously, in order to offset the ripple component in MPS current, the APF needs to output a ripple current who is an AC one, which means that the main circuit of APF is an inverter. Fig. 1 shows a schematic of a conventional parallel-APF[3].

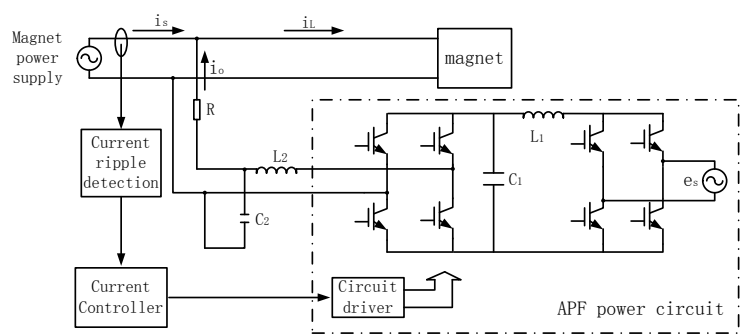

Fig.1. Schematic of a conventional parallel-APF

As shown in Fig.1, the APF mainly consists of three parts, which are the circuit, the current controller, and the current ripple detection. When the APF is working, the current ripple detection analyzes the MPS current to get the information about the ripple component, and then the APF circuit is controlled by the current controller to generate an opposite ripple current according to the detected result.

In Fig.1, $i_s$ denotes the current outputted by the MPS, $i_o$ is the compensating current from the APF, and $i_L$ is the final current through the magnet. Among them, $i_s$ can be expressed by the follow equation.

$$i_s = i_f + i_w \tag{1}$$

Where $i_f$ represents the DC component from the MPS, and $i_w$ is the ripple one. In order to make is contains only DC part, the followed formula should be satisfied.

$$i_o = -i_w \tag{2}$$

The conventional APF usually is composed of a rectifier and an inverter, between them, the rectifier converts the AC to DC, and the inverter transforms the DC into ripple current, which is AC one. These two devices make the structure of the APF complicated, meantime its volume is difficult to miniaturized, and the cost high. In addition, as the APF is connected to the magnet in parallel with the MPS, its output voltage $V_o$ is equal to the output one of MPS, which means that a voltage balance should be made between the APF and MPS. This makes the APF difficult to be applied to various MPS with different output voltage, and once the output voltage of MPS changes, the voltage of APF also needs to be altered to make a new equation.

To overcome the shortages mentioned above, a new APF topology is proposed in this paper.

## 2. Principle of a new topology

The principle of the new type of APF studied in this paper is shown in Fig.2.

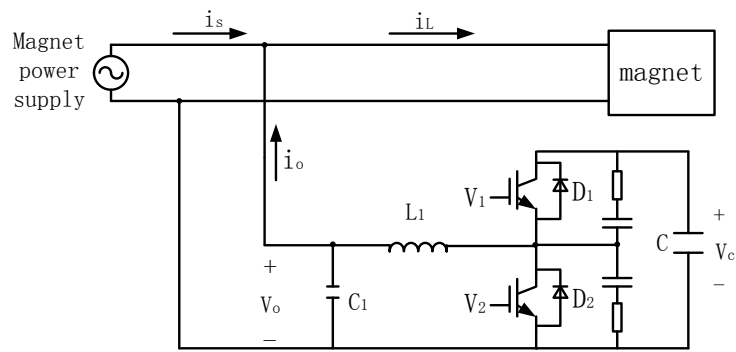

Fig. 2. Schematic of a new APF topology

The information about the amplitude, phase and frequency of ripple can be obtained by the current ripple detection. if the detected result shows $i_w>0$, then it means there is a positive ripple current in the MPS current, at this time the APF should output a negative current, which means the APF is drawing current from the MPS. At this point, make device $V_2$ work according the controller while keeping $V_1$ disconnected, then the equivalent circuit diagram of this APF can be shown in Fig.3

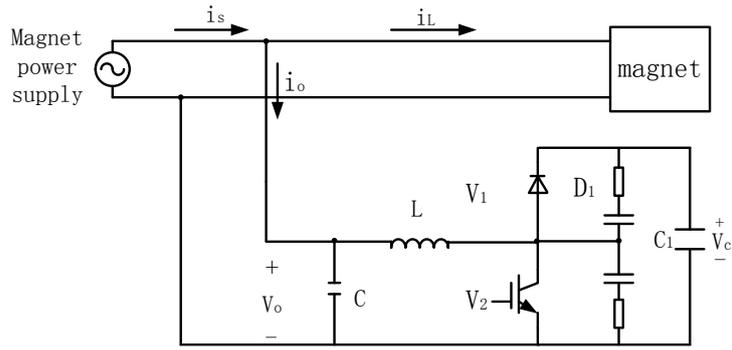

Fig.3. APF working as a Boost circuit

As shown in Fig.3, the APF works as a Boost converter in this operating state, and the APF circuit stores the ripple power from the MPS to the storage capacitor $C_1$ during $V_2$ connected. In this case, the input power voltage to the AFP is the output voltage of MPS, which is $V_o$, and the output voltage of this Boost circuit is the voltage across capacitor $C_1$, which is $V_c$.

The relationship between $V_c$ and $V_o$ can be represent as equation (3).

$$V_c = \frac{V_o}{1-D_2} \qquad (3)$$

Where $D_2$ is the duty ration of $V_2$.

Otherwise when $i_w<0$, then it predicates that the current ripple of the MPS is negative, so at this time the APF should output a positive ripple current, which means that the APF need inject a ripple current to the magnet. In this case, $V_2$ needs to be disconnected and $V_1$ works according to the controller, and then the equivalent circuit diagram can be shown in Fig.4

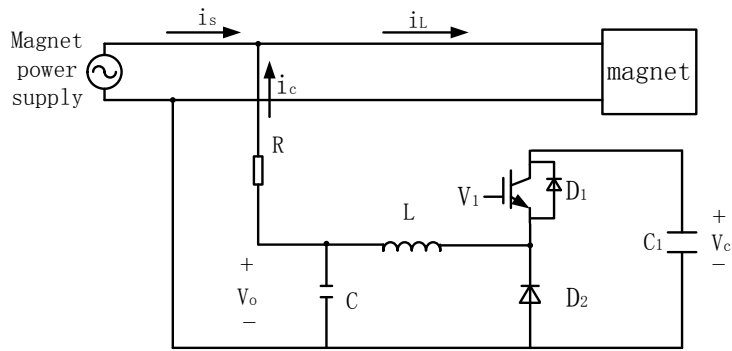

Fig.4. APF working as a Buck circuit

Obviously, the APF works as a Buck converter during this period. Its power source is the voltage across the capacitor $C_1$, and the output voltage is the MPS voltage $V_o$. The equation between the two voltages is expressed as follow.

$$V_o = V_c \times D_1 \qquad (4)$$

Where $D_1$ is the duty ration of device $V_1$.

It can be concluded from the operating process that the input and output voltages of the APF circuit is determined by the voltage of MPS, which means it is not a fixed value, but adapts automatically to the MPS voltage. This means that the APF is very adaptive, and when the MPS outputs different voltage, the APF can catch up with the changing.

The APF can be considered as a power supply when designing its controller. However, the control loop of APF is not totally identical to the normal digital power supply[4]. That is because the reference of APF circuit is not an aware value or wave as in power supply, but the detected information about the current ripple, which is variable. Fig.5 shows the diagram of the proposed APF's control loop in this paper.

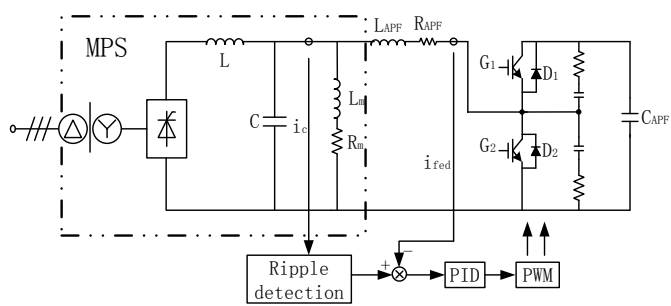

Fig.5. Schematic of APF's Control loop

## 3. APF prototype and experiments

In order to verify and examine the performance and effectiveness of the APF described above, an experimental system is built to test the work of the APF, as shown in the followed picture.

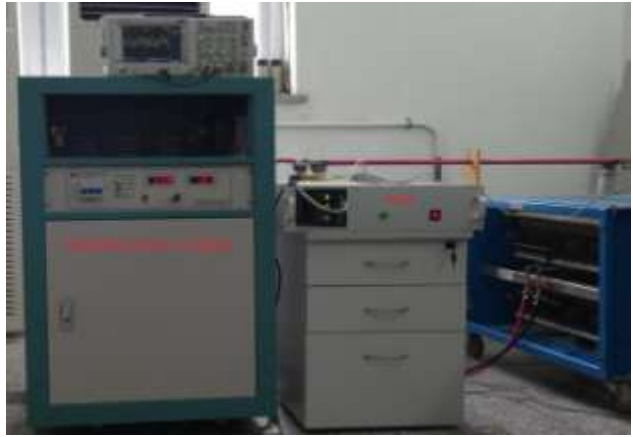

Fig.6. Experiment system of the APF

The experiment system is composed by a MPS, an APF prototype, a computer, a resistive load, and a scope. Among them, the computer is used for software debugging and parameters modulation. As described previously, the APF is connected in parallel with the MPS to the magnet, which is shown in Fig.7.

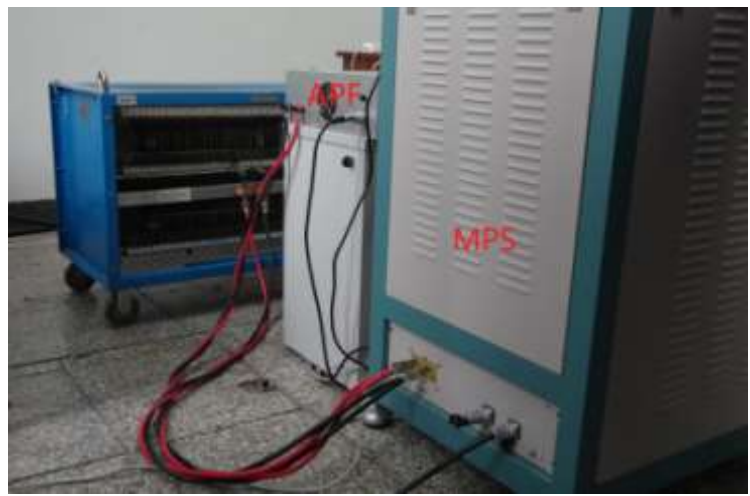

Fig.7. The connection of APF between power supply and load

In order to observe the current waveform of the load directly, a resistance is used as the load, whose current waveform is the same with voltage one. In purpose of watching the changes of current ripple, the scope is set to AC observation in this experiment. Fig.8 shows the waveform of the MPS current ripple when the MPS output current is set to 24A with the APF disconnected, and Fig.9 shows the current waveform after the APF connected.

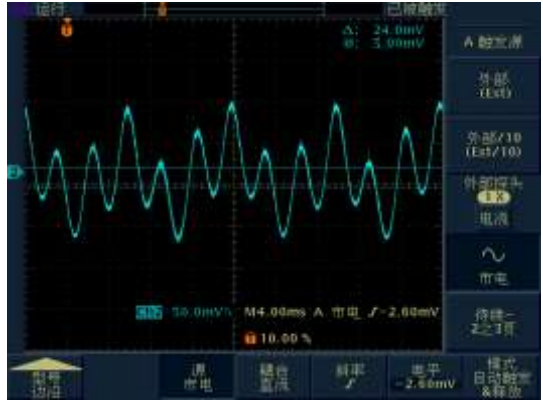

Fig.8. Waveform of current ripple when MPS working at 24A

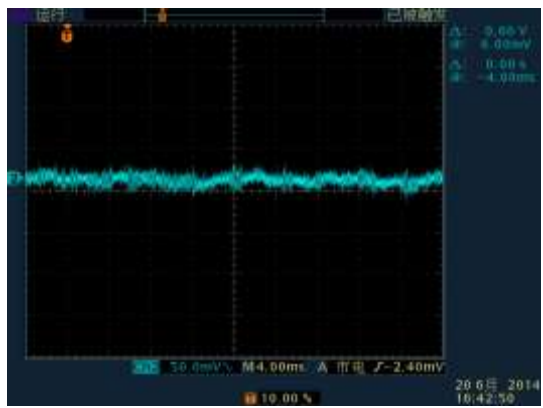

Fig.9. Waveform of current ripple With APF connected when MPS working at 24A

In Fig.8, the amplitude of current ripple is about 180mV, and the APF decrease this ripple to less than 20mV. So it can be concluded that with this APF the current ripple is decreased to about one tenth.

Turn the output current of the MPF to 40A, then the current waveforms before and after the APF's compensation are respectively shown in Fig.10 and Fig.11.

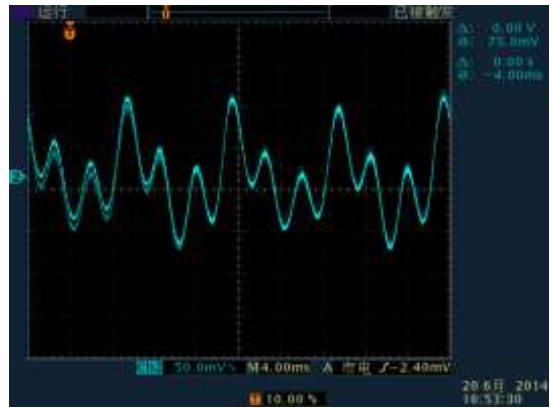

Fig.10. Waveform of current ripple when MPS working at 40A

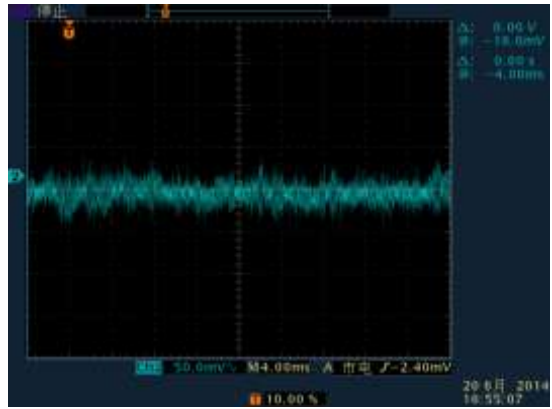

Fig.11. Waveform of current ripple With APF connected when MPS working at 40A

It can be seen in Fig.10 and Fig.11 that the APF reduces this ripple from about 200mV to 20 mV. This process not only proves the effect of APF's ripple compensation, but also testifies to its adaptation to the change of MPS voltage.

## 4. Conclusion

The follow conclusion can be concluded from the experiment mentioned above.

1. The APF in this paper can make dynamic and online compensation for the MPS current ripple. With this APF, the MPS current ripple can be reduced to a tenth of the primary.

2. The effort of the compensation effect becomes worse during the MPS current's changing, such as the current changes from 24A to 40A in the experiment. However, after the MPS current adjusts into new stability, the APF can track the changing, and ultimately realize the decrease of the MPS current ripple.

3. This APF is totally independent of the MPS system, thus it can be installed or removed conveniently according to the MPS's requirements.

4. The effect of this APF can be further improved. For this purpose, the AD samples to MPS current need much higher precision and the chosen capacitor in APF circuit should be reconsidered. In addition, the circuit of this APF needs further research and experiments.